\documentstyle[aps,prb,twocolumn,epsf,floats]{revtex}
\draft
\begin{document}
\wideabs{

\title{Spin excitation spectra 
       of integral and fractional quantum Hall systems}

\author{
   Arkadiusz W\'ojs}
\address{
   Department of Physics, 
   University of Tennessee, Knoxville, Tennessee 37996, \\
   and Institute of Physics, 
   Wroclaw University of Technology, Wroclaw 50-370, Poland}

\author{
   John J. Quinn}
\address{
   Department of Physics, 
   University of Tennessee, Knoxville, Tennessee 37996}

\maketitle

\begin{abstract}
Results are presented of detailed numerical calculations for 
the spin excitation spectra of a two-dimensional electron gas
confined in a quantum well of finite width $w$, at magnetic 
fields corresponding to the fractional and integral fillings 
of the lowest and of excited Landau levels.
Spin waves and skyrmions are identified, and their mutual 
interactions are studied at filling factors $\nu={1\over3}$, 
1, 3, and 5.
The smallest skyrmions at $\nu={1\over3}$ are equivalent to 
composite fermion charged excitons $X_{\rm CF}^\pm$.
Bose condensates of nearly noninteracting spin waves and 
Laughlin correlations between finite-size skyrmions are found 
at $\nu=1$ and ${1\over3}$ and, for $w$ larger than about
two magnetic lengths, also at $\nu=3$ and 5.
A general criterion for the occurrence of skyrmions at odd 
integral fillings and at Laughlin fractional fillings of the 
lowest Landau level is given in terms of the interaction 
pseudopotentials.
This explains the dependence of skyrmion states on $w$ observed 
in excited Landau levels.
\end{abstract}
}

\section{Introduction}

The low-energy dynamics of a two-dimensional electron gas (2DEG) 
in a strong magnetic field $B$ is determined by the particular form 
of the electron--electron ($e$--$e$) interaction in a macroscopically 
degenerate, partially filled Landau level (LL).
\cite{Prange87,Chakraborty95,DasSarma97}
In the lowest LL, the short range of this (repulsive) interaction 
results in a particular form of (Laughlin) correlations.
\cite{Laughlin83}
Namely, at each given density $\varrho$, yielding a fractional LL 
filling $\nu=\varrho\phi_0/B=2\pi\varrho\lambda^2$ ($\phi_0=hc/e$ 
is a magnetic flux quantum and $\lambda=\sqrt{\hbar c/eB}$ is the 
magnetic length), the electrons tend to avoid\cite{Haldane87,jphys} 
as much as possible (within their Hilbert space severely limited by 
the LL quantization) the pair eigenstates with the highest repulsion, 
that is with the smallest relative pair angular momentum ${\cal R}=0$, 
1, 2, \dots.
This microscopic property is responsible for variety of macroscopic, 
experimentally observable effects.
A well-know example is the fractional quantum Hall effect
\cite{Laughlin83,Tsui82} in which a finite gap for charge 
excitations at $\nu={1\over3}$, ${2\over5}$, etc., causes 
quantization of the Hall conductance at the universal values, 
$\sigma_{xy}=\nu e^2/h$.
Another example is a nonlinear dependence of the spin polarization
$\zeta$ on the magnetic field $B$ (or density) near $\nu=1$ 
(in the systems in which the Zeeman energy $E_{\rm Z}$ can be made 
sufficiently small to allow spin-excitations),\cite{Barret95,Tycko95,%
Maude96,Aifer96,Manfra96,Shukla00,Khandelwal01} connected with the 
existence of particle-like excitations carrying massive spin, 
generally called skyrmions.\cite{Skyrme61,Rajaraman82,Lee90}

It is quite remarkable that both features that determine the 
properties of the 2DEG in the quantum Hall regime -- the degenerate 
LL structure of the single-particle spectrum and the repulsive 
interaction between the particles (electrons or holes) in a 
partially filled LL -- repeat at different filling factors $\nu$.
Clearly, the situation near all odd integral $\nu=1$, 3, \dots\ 
is very similar if only the cyclotron energy $\hbar\omega_c$
is sufficiently large to prevent the inter-LL scattering 
(and cause decoupling of a partially filled excited LL from 
the rigid core of completely filled, lower LL's).
Less obviously, the Laughlin correlations in a partially filled 
lowest LL allow (at low energy) the mapping of the original 
electron system near $\nu=(2p+1)^{-1}$ onto the system of weakly 
interacting quasiparticles (QP's) with $(2p+1)$-times smaller
degeneracy $g^*$ of their quasi-LL's.\cite{Haldane83}
It is a matter of preference whether the reduced quasi-LL degeneracy 
$g^*=g/(2p+1)$ is attributed to the fractional charge $\pm e/(2p+1)$ 
of the QP's\cite{Laughlin83,Haldane83} or, as in the composite fermion 
(CF) picture,\cite{Jain89,Lopez91,Halperin93,Sitko97} to the partial 
cancellation of the magnetic field $B$ by the Chern--Simons gauge field. 
In any case, the effective QP filling factor is $\nu^*\approx1$ 
(which allows interpretation of the fractional quantum Hall effect 
of electrons as an integral quantum Hall effect\cite{Klitzing80} of 
QP's), with small residual interaction between the (charged) QP's.

The answer to the question if the low-energy spin excitations 
of all these similar systems are indeed similar (and, e.g., 
include stable skyrmions) lies in the details of this residual 
interaction within the relevant (lowest or excited, electron 
or CF) LL.
This question is not at all trivial, as it is known that the 
repulsive character of this interaction is, by itself, not 
sufficient to cause similar correlations.
For example, unlike in the lowest LL, electrons at the 
$\nu={1\over3}$ filling of an excited LL do not form 
a Laughlin ground state.\cite{acta,philmag,fivehalf}
Similarly, of the two types of Laughlin QP's, the quasiholes (QH's) 
do, but the quasielectrons (QE's) do not form a Laughlin ground 
state at a QP filling of $\nu_{\rm QP}={1\over3}$.\cite{hierarchy}
To make things more complicated, the effective 2D interaction 
in realistic quasi-2D systems depends also on the details of the 
confinement (in the $z$-direction, perpendicular to the 2D layer).
\cite{He90}

The spin excitations at $\nu=1$ have been investigated both 
experimentally\cite{Barret95,Tycko95,Maude96,Aifer96,Manfra96,%
Shukla00,Khandelwal01} and theoretically,\cite{Sondhi93,%
Fertig94,Brey95,Moon95,MacDonald96,Xie96,Cote96,Cote97,%
Abolfath97,Fertig97,Oaknin98,Melik99} and both in an extended 
2DEG and in finite-size quantum Hall droplets.\cite{Palacios94,%
Oaknin96,droplet}
Importantly, an exact mapping between the unpolarized electron 
($\uparrow$--$\downarrow$) and polarized electron--valence-hole 
($e$--$h$) systems in the lowest LL\cite{MacDonald90} makes a $\nu=1$ 
skyrmion equivalent to a charged multi-exciton\cite{x-td,Palacios96} 
($X_K^\pm$) consisting of $K$ neutral excitons bound to an electron 
or to a hole.
This allows cross-interpretation of the results obtained from 
the (experimental and theoretical) studies of spin and optical 
excitations.

Unlike in the integral quantum Hall regime, skyrmions at $\nu={1\over3}$ 
(proposed by Kamilla {\sl et al.}\cite{Kamilla96}) were only recently 
detected in a transport experiment,\cite{Leadley97,Khandelwal98} thanks 
to a sufficient reduction of the Zeeman gap $E_{\rm Z}$ by means of 
hydrostatic pressure.
The subsequent numerical calculation\cite{MacDonald98} also indicated 
the formation of small skyrmions and anti-skyrmions in finite-size
fractional quantum Hall systems.
However, the question of why similar spin excitations occur in
the $\nu=1$ electron system and the $\nu^*=1$ CF system despite
different $e$--$e$ and CF--CF interactions\cite{hierarchy} has not 
yet been answered.

The situation in the excited LL's is not yet completely understood, 
either.
First, it was predicted\cite{Jain94,Wu95} that skyrmions do not occur
at $\nu=3$, 5, etc.
This prediction was soon confirmed experimentally.\cite{Schmeller95}
Then it was found\cite{Cooper97} that a finite width of a quasi-2DEG 
stabilizes skyrmions in excited LL's.
Indeed, rapid spin depolarization around $\nu=3$ was recently observed
\cite{Song99} in a rather wide, 30~nm quantum well.
While it is clear that the finite width enters the problem of an 
isolated, excited LL only through the weakening of the $e$--$e$ 
repulsion at short range, the class of interactions for which 
skyrmions are stable has not yet been generally defined.

In this paper we compare the results of detailed numerical 
calculations of the spin excitation spectra at $\nu={1\over3}$, 
1, 3, and 5.
We identify spin waves and skyrmions in each system, and analyze 
their single particle properties as well as mutual interactions. 
The spin excitation spectra turn out to be very similar at $\nu=
{1\over3}$ and 1, and, if the layer width $w$ is sufficiently 
large, also at $\nu=3$ and 5.
In particular, we confirm that skyrmions are the lowest-energy 
charged excitations in all these systems if the Zeeman energy 
$E_{\rm Z}$ is sufficiently small.

We show that the formation of particle-like skyrmions coincides 
with the occurrence of condensed states of a macroscopic number 
of nearly noninteracting charge-neutral spin waves each with 
angular momentum $L=1$.
The interaction energy of these states is (nearly) linear in spin 
polarization $\zeta$ (and thus it remains finite at $\zeta=0$) 
giving rise to a gapless and continuous density of states.
On the other hand, the correlations between charged (particle-like) 
skyrmions are expected to be of Laughlin type,\cite{Laughlin83} 
meaning that the skyrmion pair eigenstates with the smallest 
relative pair angular momentum ${\cal R}$ are maximally avoided.
\cite{Haldane87,jphys,acta,philmag,fivehalf}
This can be rephrased in terms of an effective spatial isolation 
of skyrmions from one another, and the absence of high-energy 
skyrmion--skyrmion collisions (at low temperature).
\cite{x-fqhe,x-cf,x-tb}

We also determine a general criterion for the occurrence of 
skyrmions in a system of spin-${1\over2}$ particles half-filling 
a spin-degenerate shell (e.g., in a system of electrons or 
Laughlin QP's at $\nu=1$ and an arbitrary layer width $w$).
We find that the particle--particle interaction pseudopotential
\cite{Haldane87} $V_{\rm ee}({\cal R})$ must:
(i) be strongly repulsive (superharmonic\cite{acta,philmag}) at 
${\cal R}=0$ to cause decoupling of the many-body states that 
avoid having ${\cal R}=0$ pairs from all other states (skyrmions 
are exact eigenstates of the ideal short-range repulsion with 
$V_{\rm ee}(0)=\infty$), and
(ii) decrease sufficiently quickly with increasing ${\cal R}$ 
between ${\cal R}=1$ and 3, ${\cal R}=3$ and 5, etc.\ to make 
the skyrmion energy decrease with increasing the topological 
charge $K$ (size) and, in particular, bring the skyrmion energy 
band below the energy of the spin-polarized QP state.
For systems with broken particle($\uparrow$)--hole($\downarrow$) 
symmetry (e.g., at $\nu={1\over3}$), the latter condition must 
be rephrased in terms of the particle--hole pseudopotential 
$V_{\rm eh}$ which (iii) must increase monotonically as a function 
of wave vector $k$.

The above criterion allows prediction and explanation of the 
occurrence or absence of skyrmions (and the resulting type of 
dependence of spin polarization $\zeta$ of the 2DEG on density 
or magnetic field) at an arbitrary filling factor $\nu$, layer 
width $w$, density profile across the layer $\varrho(z)$, etc.\ 
-- based on the analysis of the involved $e$--$e$, $e$--$h$, 
QE--QE, QH--QH, or QE--QH interaction pseudopotentials.
This criterion is somewhat analogous to the one for Laughlin 
correlations in a partially filled shell,
\cite{jphys,acta,philmag,fivehalf} which explained compressibility
\cite{hierarchy} of the spin-polarized $\nu={4\over11}$ state 
($\nu={1\over3}$ state of Laughlin QE's) and incompressibility
\cite{qer} of the partially unpolarized state at the same 
$\nu={4\over11}$ ($\nu={1\over3}$ state of reversed-spin QE's
\cite{Chakraborty86,Rezayi87}).

\section{Model}

The model used here is essentially that of Ref.~\onlinecite{qer}, 
except that in the present calculation we do not include valence-band 
holes.
A finite system of $N$ electrons is confined on Haldane sphere
\cite{Haldane83} of radius $R$.
The radial magnetic field $B$ is due to Dirac monopole, whose 
strength $2Q$ is defined in units of flux quantum $\phi_0$, 
so that $4\pi R^2B=2Q\phi_0$ and $R^2=Q\lambda^2$.
The single-electron states\cite{Haldane83,Wu76,Fano86} are the 
eigenstates of magnitude and projection of angular momentum 
($l$ and $m$) and of the spin projection ($\sigma$), and form 
$g$-fold ($g=2l+1$) degenerate LL's labeled by $n=l-Q=0$, 1, 
2, \dots.

The cyclotron energy $\hbar\omega_c\propto B$ is assumed much 
larger than the Coulomb interaction energy scale $E_{\rm C}=e^2/
\lambda\propto\sqrt{B}$, so that excitations between LL's can 
be neglected, and only one, isolated $n$th LL need be considered.
On the other hand, no assumption is made about the Zeeman spin 
splitting $E_{\rm Z}$, which in GaAs samples can be made arbitrarily 
small by applying hydrostatic pressure.
The ratio $\eta=E_{\rm Z}/E_{\rm C}$ of energy scales associated 
with spin and charge excitations (within an isolated LL) is 
a small free parameter of the model.

Unless it is much larger than $\lambda$, finite layer width $w$ 
enters the problem by only modifying the quasi-2D interaction 
pseudopotential $V({\cal R})$.
In the lowest LL, the effects due to a finite $w$ can be 
adequately modeled by merely reducing the Coulomb energy scale 
$E_{\rm C}$ by a factor $\xi_w<1$ compared to the ideal $w=0$ 
case ($\xi_w\sim{1\over2}$ in typical samples).
However, the situation is very different in the excited LL's,
where even if $w$ is sufficiently small that the inter-subband 
mixing can be neglected, it affects not only the overall energy 
scale\cite{Cooper97} but the shape of $V({\cal R})$, and thus the 
electron correlations\cite{jphys,acta,philmag,fivehalf} as well.
To address this problem, we have included finite $w$ in the
calculation for excited LL's by using the quasi-2D Coulomb 
$e$--$e$ potential $V_d(r)=e^2/\sqrt{r^2+d^2}$, which is known
\cite{He90,x-tb} to reproduce well the pseudopotential 
$V({\cal R})$ for $w\approx5d$.
Here, $w$ is the effective layer width obtained from fitting 
the actual lowest-subband density profile $\varrho(z)$ with
$\cos^2(\pi z/w)$.
In typical GaAs wells $w$ is larger than the well width by 3--3.5~nm 
due to a finite barrier height.

The Hamiltonian of interacting electrons confined to the $n$th 
LL can be written as
\begin{equation}
   H=
   \sum
   c_{m_1\sigma}^\dagger
   c_{m_2\sigma'}^\dagger
   c_{m_3\sigma'}
   c_{m_4\sigma}
   \left<m_1m_2|V|m_3m_4\right>,
\label{eqham}
\end{equation}
where $c_{m\sigma}^\dagger$ and $c_{m\sigma}$ are the electron 
creation and annihilation operators, and the interaction matrix 
elements are calculated for the potential $V_d(r)$ and they are 
connected with the pseudopotential $V({\cal R})$ through the 
Clebsch--Gordan coefficients.
Hamiltonian $H$ is diagonalized in the basis of $N$-electron 
Slater determinants
\begin{equation}
   \left|m_1\sigma_1 \dots m_N\sigma_N\right>=
   c_{m_1\sigma_1}^\dagger 
   \dots 
   c_{m_N\sigma_N}^\dagger
   \left|{\rm vac}\right>,
\label{eqbasee}
\end{equation}
where $\left|{\rm vac}\right>$ stands for the vacuum state.
While using basis (\ref{eqbasee}) allows automatic resolution 
of two good many-body quantum numbers, projection of spin 
($S_z=\sum\sigma_i$) and of angular momentum ($L_z=\sum m_i$), 
the length of spin ($S$) and of angular momentum ($L$) are 
resolved numerically in the diagonalization of each $(S_z,L_z)$ 
Hilbert subspace.
The results obtained on Haldane sphere are easily converted to 
the planar geometry, where $L$ and $L_z$ are appropriately
\cite{Avron78,geometry} replaced by the total and center-of-mass 
angular momentum projections, $M$ and $M_{\rm CM}$.

\section{Integral quantum Hall regime (lowest Landau level)}

\subsection{Exactly filled level}
\begin{figure}[t]
\epsfxsize=3.40in
\epsffile{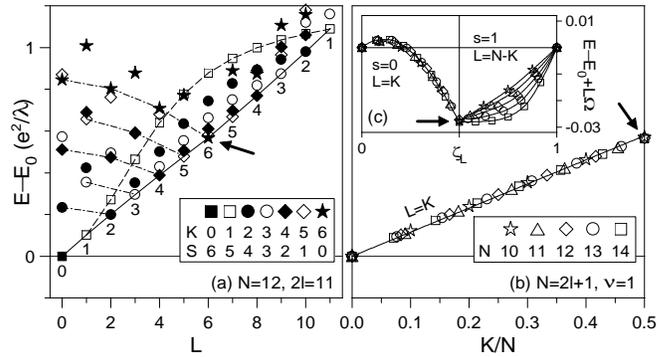}
\caption{
   (a) 
   The energy spectrum 
   (Coulomb energy $E$ versus angular momentum $L$) 
   of the system of $N=12$ electrons 
   in the lowest ($n=0$) LL of degeneracy $g=2l+1=12$ 
   calculated on Haldane sphere.
   $S$ is the total spin, 
   and $K={1\over2}N-S$ is the number of reversed spins.
   Dashed line: single spin wave;
   dash-dot lines: states containing equal numbers of $L=1$ 
      spin waves;
   solid line: condensates of noninteracting $L=1$ spin waves.
   (b)
   The energy dispersion 
   (energy $E$ versus spin polarization $\zeta=K/N$)
   for the spin-wave condensate states at $\nu=1$, 
   calculated for $N\le14$ electrons 
   exactly filling their lowest LL ($N=2l+1$).
   (c)
   The dispersion of (b), but with inclusion of a lateral 
   confinement term $L\Omega$ and plotted as a function of
   normalized angular momentum $\zeta_L$.
   $\lambda$ is the magnetic length and 
   arrows mark the same state in each frame.}
\label{fig01}
\end{figure}

Let us begin the discussion of spin excitations of the 2DEG in the 
integral quantum Hall regime with the energy spectrum at precisely 
$\nu=1$.
Similar spectra were previously studied in the context of finite-size
quantum Hall droplets.\cite{Oaknin96,droplet}
As an example, in Fig.~\ref{fig01}(a) we show the finite-size spectrum 
for $N=12$ electrons on Haldane sphere with LL degeneracy $g=2l+1=N$.
In this and all other spectra in the paper, the excitation energy 
$E-E_0$ is given in the units of $E_{\rm C}=e^2/\lambda$, measured 
from the energy $E_0$ of the lowest maximally spin-polarized state 
(here, the $\nu=1$ ground state), excludes the Zeeman energy 
$E_{\rm Z}$, and is plotted as a function of total angular momentum 
$L$ (for waves, related to longitudinal wave vector $k$ by $L=kR=
k\lambda\sqrt{Q}$).
Different symbols mark multiplets of different total spin $S$ 
and only the lowest state is shown at each $L$ and $S$.
The labels show the number $K$ of reversed spins (relative to the
maximally polarized state) defined as $K={1\over2}N'-S$, where 
$N'={\rm min}(N,g-N)$.

It is well-known that even in the absence of the Zeeman energy gap
($E_{\rm Z}=0$) the ground state (GS) of the 2DEG at $\nu=1$ is 
spin-polarized (ferromagnetic) and translationally invariant 
(hence, nondegenerate).
At the exact half-filling of a spin-degenerate LL, this GS is the 
only state with zero projection onto the most strongly repulsive 
$e$--$e$ pair eigenstate at ${\cal R}=0$.
Indeed, Fig.~\ref{fig01}(a) shows a GS at $K=L=0$.
Because of the complete filling of the spin-$\downarrow$ $n=0$ LL 
in the ferromagnetic GS, the only excitations below the cyclotron 
gap $\hbar\omega_c$ are the spin waves (SW's) with $K=1$ and $L=1$,
2, 3, \dots.
A single SW consists of a vacancy (hole) in the spin-$\downarrow$ 
level and an electron in the spin-$\uparrow$ level, and is also 
called a spin-exciton.
Its dispersion curve is given by\cite{Kallin84} $E_{\rm SW}(k)=E_0
+E_{\rm Z}+E_{\rm C}\sqrt{\pi/2}[1-\exp(-\kappa^2)I_0(\kappa^2)]$,
where $\kappa=k\lambda/2$ and $I_0$ is the modified Bessel function, 
and it is shown in Fig.~\ref{fig01}(a) with a dashed line.

Interestingly, Fig.~\ref{fig01}(a) shows that a single SW is not 
the lowest-energy excitation at $L>1$.
Instead, the lowest excitations form a band with $K=L$ and energy 
nearly linear in $K$ (solid line).
This (nearly) linear dependence $E(K)$ can be interpreted as the 
(nearly) perfect decoupling of SW's each with $L=1$, earlier pointed 
out by Oaknin {\sl et al.}\cite{Oaknin96}.
Denoting the energy of one SW by $\varepsilon_{\rm SW}$ allows 
us to approximate the lowest state with $K$ reversed spins by 
$E_{\rm SW}(K)=E_0+K\varepsilon_{\rm SW}$.
Note that a pair of SW's each with $L=1$ can be in either parallel 
or anti-parallel state, with $K=2$ and pair angular momentum 
$L=2$ or 0, respectively.
It follows from Fig.~\ref{fig01}(a) that the $L=2$ state containing
a pair of SW's is noninteracting, while the $L=0$ state of such 
a pair is repulsive (the SW--SW repulsion energy at $L=0$ decreases 
as a function of system size, and for $N=12$ it is $\approx0.03\,
E_{\rm C}$).
Similarly, the states containing the number $K$ of SW's each with 
$L=1$ (see dotted lines) can have total angular momentum $L=K$, 
$K-2$, $K-4$, \dots.
However, only the $L=K$ state does not contain any SW--SW pairs 
with $L=0$, and thus only this state has energy $E=E_{\rm SW}(K)$, 
while all others have $E>E_{\rm SW}(K)$.
The $L=K$ low-energy excitations of a 2DEG at $\nu=1$ are uniquely 
ordered states of noninteracting SW's each with $L=1$ (Bose--Einstein 
condensates) and will be denoted $W_K$.

The exact mapping\cite{MacDonald90} between the two-spin electron 
system ($\uparrow$--$\downarrow$) and the electron--valence-hole 
($e$--$h$) system in the lowest LL allows the expression of the 
above statement in terms of the interaction between interband 
magneto-excitons; namely, the $L=1$ excitons in the lowest LL do 
not interact with one another and they condense into correlated 
$L=K$ states.
Although similar, this symmetry is independent from the well-known
``hidden symmetry''\cite{Dzyubenko83} and the decoupling of $L=0$ 
excitons.\cite{MacDonald90,x-fqhe,x-cf,x-tb} 
Its consequence will be also different because, in contrast to 
the radiative $L=0$ excitons, the $L=1$ excitons are long-lived.

To determine the energy spectrum of an infinite 2DEG we have 
compared data obtained for different electron numbers, $N\le14$.
As shown in Fig.~\ref{fig01}(b), it turns out that the energy 
$E_W$ of $W_K$ excitations is an almost linear function of the 
``relative'' spin polarization, $\zeta=K/N$.
Namely, $E_W=E_0+u\zeta$, with the slope that we estimate as 
$u\approx1.15\,E_{\rm C}$.
This fact reflects the extended character of the SW's and has 
a couple of obvious consequences for the thermodynamic limit 
of $N\rightarrow\infty$:
(i) 
For any $E_{\rm Z}\ne0$, the interaction energy $E_W(K)-E_0\propto 
K/N$ of each correlated $W_K$ state is negligible compared to its 
total Zeeman energy, $KE_{\rm Z}$.
(ii) 
The gap for spin excitations is precisely the Zeeman gap $E_{\rm Z}$; 
if this gap can be closed  (e.g., by applying hydrostatic pressure), 
the $\nu=1$ ferromagnetic GS becomes gapless, the density of states 
for the $W_K$ excitations becomes continuous, and a macroscopic number 
$\sim Nk_{\rm B}T/u$ of noninteracting SW's occur at an arbitrarily 
small finite temperature.

To remove the dominant linear term from $E_W(\zeta)$ and study the 
small nonlinear correction, in Fig.~\ref{fig01}(c) we plot the energy 
spectrum shifted by an additional term $L\Omega$.
Physically, this term describes a harmonic lateral confinement applied 
to a finite-size $\nu=1$ droplet.\cite{Palacios94,Oaknin96,droplet}
The confinement strength $\Omega$ is chosen so that the $\nu=1$ GS 
is degenerate with the next ``compact droplet'' state at $K=1$ and 
$L=N-1$ (compact-droplet eigenstates\cite{Palacios94,Oaknin96,droplet}  
$C_K$ with $K\ge1$ are created by 
$\prod_{i=0}^{K-1}c_{l-i,\uparrow}^\dagger c_{-l+i\downarrow}$ 
acting on 
$C_0=\prod_{m=-l}^l c_{m,\downarrow}^\dagger \left|{\rm vac}\right>$,
i.e.\ on the $\nu=1$ GS).
To compare data for different $N$, the energy is plotted as a function 
of normalized angular momentum, $\zeta_L=(L-s^2)/(N-2s)$, where $s$ is 
equal to $K$ of the nearest $C_K$ state, so that $\zeta_L$ increases 
with $L$ and equals $K$ in each $C_K$ state (note that $\zeta_L=\zeta$ 
for $\zeta_L\le{1\over2}$).

Clearly, the nonlinear correction to $E_W-E_0$ is also a well-defined 
function of $\zeta$ or $\zeta_L$ with a minimum at $\zeta_L={1\over2}$, 
i.e. in the highly correlated singlet ($S=0$) state marked with arrows.
If both $E_{\rm Z}$ and $N$ are sufficiently small, the energy of 
this singlet state, $E_W({1\over2})-E_0+{1\over2}N\Omega
\approx-0.03\,E_{\rm C}+{1\over2}NE_{\rm Z}$ can remain negative.
Then, the sublinear behavior of $E(\zeta_L)$ between $\zeta_L=0$ 
and ${1\over2}$ implies an abrupt transition from the $\nu=1$ 
ferromagnetic GS to the $\zeta_L={1\over2}$ singlet GS (skipping 
intermediate spins $S$) of a finite-size quantum Hall droplet 
as a function of $\Omega$.\cite{Oaknin96,droplet}
On the other hand, the superlinear behavior of $E(\zeta_L)$ between 
$\zeta_L={1\over2}$ and 1 implies that a further decrease of $\Omega$ 
drives a quantum Hall droplet from the $S=0$ state to the $S=N-1$ 
compact state through a series of GS's with all intermediate spins.

\subsection{Additional particle or hole in filled level}
What truly ignites the abrupt depolarization of a finite $\nu=1$ 
droplet under the variation of the lateral confinement is the 
insertion of an additional reversed-spin particle (from the edge) 
into the bulk of the droplet.
This effect is a consequence of a more general phenomenon -- the 
occurrence of particle-like charged excitations with macroscopic 
spin called skyrmions.\cite{Skyrme61,Rajaraman82,Lee90}
Let us consider a (possibly infinite) system of spin-${1\over2}$ 
fermions half-filling ($\nu=1$) a degenerate ($E_{\rm Z}=0$) shell 
of single-particle states and forming a ferromagnetic GS in 
accordance with a standard atomic Hund's rule.
Depending on details of the interaction pseudopotential, addition
of a single particle or hole to such system may or may not cause
its complete depolarization, that is a transition to a highly 
correlated singlet GS in which half of the electrons flipped 
their spins.
Examples of systems in which such depolarization does or does 
not occur are electrons in the lowest LL or in an atomic shell, 
respectively.

\begin{figure}[t]
\epsfxsize=3.40in
\epsffile{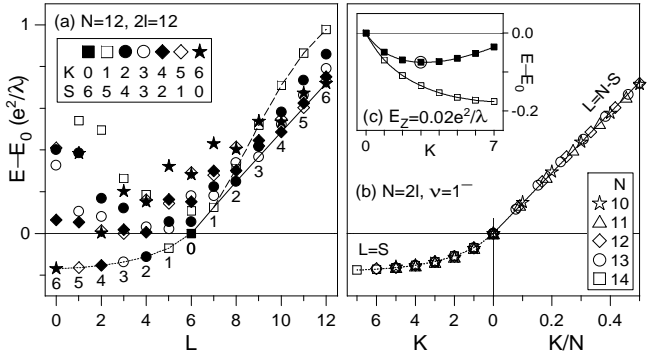}
\caption{
   (a) 
   Same as Fig.~\ref{fig01}(a) but for $g=2l+1=13$.
   Dashed line: single spin wave in the presence of a hole;
   solid line: spin-wave condensates in the presence of a hole;
   dotted line: skyrmions.
   (b)
   Same as Fig.~\ref{fig01}(b) but for $N=2l$, 
   i.e. one hole in the $\nu=1$ state.
   The skyrmion and spin-wave condensate energies are plotted 
   as a function of numbered of reversed spins $K$ and spin 
   polarization $\zeta=K/N$, respectively.
   (c)
   Open and full symbols: skyrmion dispersion of (b) for $N=14$
   without and with Zeeman energy $KE_{\rm Z}$.}
\label{fig02}
\end{figure}

The effect of removing an electron from a $\nu=1$ GS is presented 
in Fig.~\ref{fig02}(a), showing the energy spectrum of $N=12$ electrons 
at $2l=12$ (i.e., $N$ smaller by one than the LL degeneracy $g=2l+1$).
Similar spectra were first analyzed by Xie and He.\cite{Xie96}
Because of the exact particle-hole symmetry in the lowest LL, 
this system can be viewed as containing either one hole or 
one extra electron in a $\nu=1$ GS of 13 electrons, and will
be labeled as $\nu=1^\pm$.
Clearly, while the right-hand part of the spectrum (the single 
SW and $L=N-S$ bands at $L\ge{1\over2}N$) resembles that of 
Fig.~\ref{fig01}(a), a new band of states with $L=S$ appears at 
$0\le L<{1\over2}N$, at the energy below the lowest $K=0$ state.
These are the skyrmion states of topological charge $K$, denoted 
here by $S_K$ and first identified in nuclear physics by Skyrme
\cite{Skyrme61} and in the fractional quantum Hall systems by 
Sondhi {\sl et al.}\cite{Sondhi93} 
In these states, a number $K=1$, 2, \dots of SW's (spin-excitons) 
binds to a hole (strictly speaking, skyrmions consist of SW's 
bound to an electron, and the particle-hole symmetric state 
consisting of SW's and a hole is called an anti-skyrmion).
In the $e$--$h$ picture, the $S_K$ states map\cite{MacDonald90}
onto the charged multi-excitons\cite{x-td,Palacios96} $X_K^\pm$, 
consisting of $K$ $e$--$h$ pairs bound to an extra $e$ or $h$.

To understand the energy spectrum of an infinite 2DEG at $\nu=1^\pm$,
in Fig.~\ref{fig02}(b) we compare data obtained for different $N$.
Similarly to the $L=K$ band at $\nu=1$, the energy of $L=N-S$ states
at $\nu=1^\pm$ turns out to be a nearly linear function of the spin 
polarization, $E_W'=E_0+u'\zeta$, only with a slightly increased slope, 
$u'=u+v$ with $v\approx0.3\,E_{\rm C}$.
Since the angular momentum $L=N-S$ can be obtained by adding $L=K$ 
for the $W_K$ condensate and $l={1\over2}N$ for the single hole 
(assuming their parallel orientation), energy $v/N$ can be interpreted 
as the repulsion between a finite-size hole and (oriented ``parallel'' 
to it) one extended SW.

In view of their charged multi-exciton interpretation in the $e$--$h$ 
picture, it is not surprising that the relevant quantum number to
label the $L=S$ skyrmion excitations $S_K$ in an infinite system is 
$K$ (and not $\zeta=K/N$ appropriate for $W_K$ condensates).
Indeed, Fig.~\ref{fig02}(b) shows that the excitation energy 
$E_S-E_0$ of the $S_K$ states is a function of $K$ (rather 
than of $\zeta$), with the discrete series of values of $E_S(K)$ 
quickly converging as $N$ is increased (careful extrapolation to 
$N\rightarrow\infty$ gives\cite{Palacios96} $E_S(K)-E_0=-0.0529$, 
$-0.0828$, $-0.1018$, \dots, $-{1\over4}\sqrt{\pi/2}=-0.3133$ for 
$K=1$, 2, 3, \dots, $\infty$; all in units of $E_{\rm C}$).
The fact that $E_S(K)-E_0<0$ means that in the $S_K$ state 
the attraction between the hole (or electron) and $K$ SW's overcomes 
the creation energy of these SW's, and the ferromagnetic state
with $K=0$ may become unstable.
Most importantly, as shown in Fig.~\ref{fig02}(c), if $E_{\rm Z}$ 
is nonzero but smaller than $E_0-E_S(K)$, then {\em regardless 
of its actual value} a particle-like GS will occur with an excitation 
gap that is much smaller than the gap at precisely $\nu=1$ (which 
is $E_{\rm Z}$).
In other words, introduction of additional electrons (or holes) 
to the incompressible $\nu=1$ GS with a gap $E_{\rm Z}$ will cause 
significant reduction of the gap for spin excitations, and the 
objects that are able to reverse spin at low energy (much below 
$E_{\rm Z}$) are finite-size, charged particles (skyrmions) that 
move in the underlying $\nu=1$ fluid on electron-like cyclotron 
orbits. 
The ability of (mobile) skyrmions to increase and decrease spin 
at an energy cost that is small compared to and largely independent 
of $E_{\rm Z}$ (all in contrast to the $\nu=1$ state) causes critical 
magnetic field dependence of the spin relaxation rate for magnetic 
particles interacting with the 2DEG, such as ions, nuclei, or 
charged excitons.

While the decoupling of SW's follows from the linear dependence 
of $E_W$ on $K$ (or $E_W'$ on $K$, for the SW's in the presence 
of an extra electron or a hole), the nature of interaction between 
skyrmions follows from an earlier study of the $e$--$h$ complexes.
It was shown\cite{x-fqhe,x-cf} that the interaction between any 
pair of charged excitons $X_K^\pm$ ($X_0^\pm$ means an electron 
or a hole) with equal charge ($\pm e$) but possibly different sizes 
($K\ne K'$) is repulsive and similar to the $e$--$e$ interaction.
In particular, all $K$--$K'$ repulsion pseudopotentials (defined
\cite{Haldane87} as the pair interaction energy as a function of 
pair angular momentum) have short range, implying\cite{acta,philmag}
that an $X_K^\pm$ particle will have Laughlin correlations
\cite{Laughlin83} with all other $X_{K'}^\pm$ particles in the system.
Laughlin correlations are described by an appropriate Jastrow prefactor
in the many-body wave function and mean the tendency to avoid the pair 
states with maximum repulsion (minimum average separation).
This implies that (at low temperature) an $X_K^-$ does not 
undergo high-energy collisions with any other $X_{K'}^-$ charges.
Since the same must hold for skyrmions, we conclude that regardless 
of their size ($K$) or density, the skyrmions will be effectively 
isolated from one another.
This makes finite size skyrmions well-defined quasiparticles, 
virtually unperturbed by the skyrmion--skyrmion scattering, 
and excludes many-skyrmion effects from a possible spin coupling 
of a 2DEG to magnetic particles.

\section{Fractional quantum Hall regime}

Let us now, following Kamilla {\sl et al.}\cite{Kamilla96} and 
MacDonald and Palacios,\cite{MacDonald98} turn to the question 
if the spin excitations analogous to the spin-wave condensates 
and $S_K$ skyrmion particles described in the preceding section 
might also occur in the fractional quantum Hall regime.
On one hand, it is known that Laughlin correlations\cite{Laughlin83} 
in an electron system near $\nu=(2p+1)^{-1}$ (where $p$ is an integer) 
allow the mapping\cite{Jain89,Lopez91,Halperin93} of the low-energy 
states onto the noninteracting CF states with an effective filing 
factor $\nu^*\approx1$.
This mapping is done by replacing the electron LL degeneracy $g$ 
by $g^*=g-2p(N-1)$, which can be interpreted as attachment of $2p$ 
magnetic flux quanta to each electron.
On a sphere,\cite{jphys} this replaces $2l=2Q\approx(2p+1)(N-1)$ by 
$2l^*=2Q^*\approx N-1$.
On the other hand, it is the specific form of the interactions 
between the reversed-spin electrons and holes at $\nu=1$ that causes 
occurrence of $W_K$ and $S_K$ excitations, and the interaction between 
these excitations in an electron system is quite different from the 
residual interaction between CF's.
A well-known example demonstrating that the analogy between the 
electron and CF systems does not always work because of different 
interactions is the postulate of similar Laughlin correlations at 
the $\nu=(2p+1)^{-1}$ fillings of electron or CF LL's, giving rise 
to Haldane hierarchy of incompressible fractional quantum Hall 
states.\cite{Haldane83} 
For example, the $\nu={1\over3}$ states of the vacancies in the 
spin-$\downarrow$ $n=0$ CF LL (Laughlin QH's) and of particles 
in the spin-$\downarrow$ $n=1$ CF LL (Laughlin QE's) or in the
spin-$\uparrow$ $n=0$ CF LL (reversed-spin quasielectrons,
\cite{Chakraborty86,Rezayi87} QE$_{\rm R}$) correspond to the 
polarized strongly incompressible $\nu={2\over5}$ and compressible 
$\nu={4\over11}$ states,\cite{hierarchy} and to the partially 
unpolarized weakly incompressible\cite{qer} $\nu={4\over11}$ state, 
respectively.

\begin{figure}[t]
\epsfxsize=3.40in
\epsffile{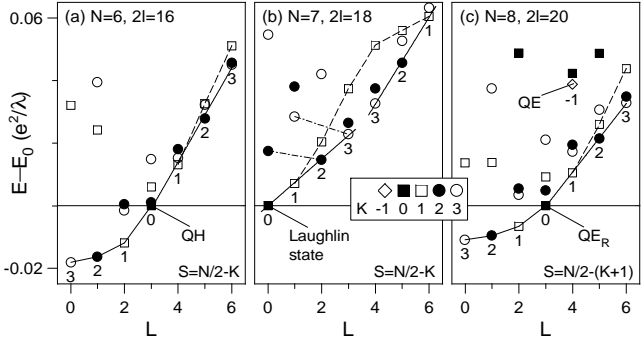}
\caption{
   Same as Figs.~\ref{fig01}(a) and \ref{fig02}(a) 
   but for the LL degeneracy $g=2l+1$ 
   equal (b), greater by one (a), or smaller by one (c) 
   to/from the value corresponding to $\nu={1\over3}$.}
\label{fig03}
\end{figure}

The examples of energy spectra at $\nu\approx{1\over3}$ are shown
in Fig.~\ref{fig03}.
The values of $N$ and $2l$ are chosen so that $g^*=7$ in each 
frame and the ``reference'' state with $K=0$ and $E=E_0$ is one 
QH ($g^*=N+1$ and $\nu^*=1^-$), Laughlin state ($g^*=N$ and 
$\nu^*=1$), and one QE$_{\rm R}$ ($g^*=N-1$ and $\nu^*=1^+$) 
in Figs.~\ref{fig03}(a), (b), and (c), respectively.
Some of the energies for smaller values of $N$ and/or $K$ have
been recently obtained by MacDonald and Palacios.\cite{MacDonald98}

Clearly, the SW dispersion $E_{\rm SW}(K)$, the linear $E_W(K)$ 
band, as well as the $E_S(K)<0$ band are all present in the spectra, 
in analogy to Figs.~\ref{fig01}(a) and \ref{fig02}(a).
What is visibly different from the $\nu=1$ spectra is a smaller 
energy scale (predominantly due to a fractional charge of involved 
QH, QE, and QE$_{\rm R}$ quasiparticles), and a discrepancy between
the $\nu=1^-$ and $1^+$ spectra.
The latter reveals only approximate particle--hole symmetry within 
the lowest CF LL (it was shown earlier\cite{qer} that the QH--QH 
and QE$_{\rm R}$--QE$_{\rm R}$ interactions are quite different).
This implies broken symmetry between the skyrmion and anti-skyrmion 
states,\cite{MacDonald98} in contrast to the $\nu=1$ case.

\begin{figure}[t]
\epsfxsize=3.40in
\epsffile{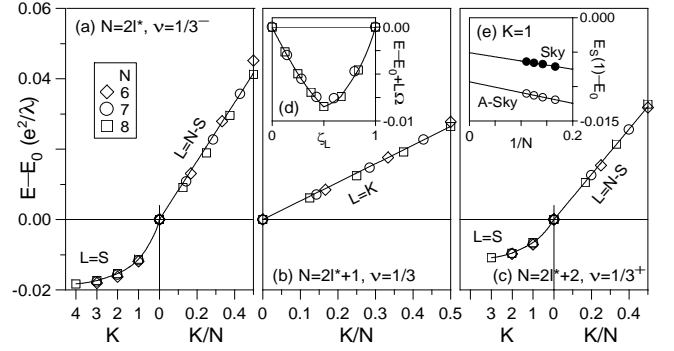}
\caption{
   Same as Figs.~\ref{fig01}(b) and \ref{fig02}(b) 
   but for the LL degeneracy $g=2l+1$ 
   equal (b), greater by one (a), or smaller by one (c) 
   to/from the value corresponding to $\nu={1\over3}$.
   Insets:
   (d) Same as Fig.~\ref{fig01}(c) but for $\nu={1\over3}$.
   (e) Energies $E_S(1)$ of skyrmions (at $\nu={1\over3}^+$)
       and anti-skyrmions (at $\nu={1\over3}^-$) with $K=1$
       as a function of inverse electron number.}
\label{fig04}
\end{figure}

Let us analyze the spectra in Fig.~\ref{fig03} in more detail.
The charge excitations of the Laughlin $\nu={1\over3}$ GS in 
Fig.~\ref{fig03}(b) are all above the magneto-roton gap 
($E-E_0\approx0.081\,E_{\rm C}$ for $N=7$) and are not shown.
The spin excitations include a single SW (an excitonic bound 
state of a QH--QE$_{\rm R}$ pair) marked by a dashed line.
Below a single SW, there is a linear band of $W_K$ states with
$L=K$ which, in analogy to the $\nu=1$ spectra, can be interpreted 
as Bose condensates of $K$ noninteracting SW's each with $L=1$ 
and all having parallel angular momenta.
As shown in Fig.~\ref{fig04}(b), similarly as at $\nu=1$, the
energy of $W_K$ states is a nearly linear function of $\zeta$,
$E_W=E_0+u\zeta$, with $u=0.053\,E_{\rm C}$ (much less than 
${1\over9}$ times the value for $\nu=1$, that would only take 
into account the smaller charge of $\pm{1\over3}e$ for QH's 
and QE$_{\rm R}$'s).
As at $\nu=1$, $E_W(K)-E_0\propto K/N$ means no gap and 
a continuous density of states if the Zeeman gap can be closed.
The comparison of Figs.~\ref{fig04}(e) and \ref{fig01}(c) shows
that applying lateral confinement can also force the spin-polarized 
finite-size $\nu={1\over3}$ fractional quantum Hall droplet to 
undergo transition to the spin-singlet ($S=0$) correlated GS at 
$L={1\over2}N$, just as it was at $\nu=1$.

The linear bands at $L=N-S$ found in Fig.~\ref{fig02} occur also at 
$\nu={1\over3}^\pm$ in Figs.~\ref{fig03}(a) and (c).
By analogy, these states correspond to a number $K$ of SW's each with 
$L=1$, coherently created in the presence of a QH or QE$_{\rm R}$.
The QH--SW and QE$_{\rm R}$--SW interaction constants $v$, obtained 
from the $E_W'=E_0+(u+v)\zeta$ fits as shown in 
Figs.~\ref{fig04}(a) and (c), are remarkably different, $0.030\,
E_{\rm C}$ and $0.011\,E_{\rm C}$, respectively.

The discrete skyrmion and anti-skyrmion bands at $L=S$ in 
Figs.~\ref{fig03}(a) and (c) also resemble their $\nu=1$ counterpart 
in Fig.~\ref{fig01}(a), and the energies $E_S(K)$ all seem 
to converge when $N\rightarrow\infty$.
For example, in Fig.~\ref{fig04}(e) the linear extrapolation of the 
energies of skyrmions (Sky) and anti-skyrmions (A-Sky) with $K=1$, 
obtained for $N\le9$, gives $E_S(1)-E_0=-0.0050\,E_{\rm C}$ 
and $-0.0093\,E_{\rm C}$, respectively (compare with $-0.0529\,
E_{\rm C}$ for $\nu=1$).
These are the critical values of the Zeeman energy $E_{\rm Z}$, 
below which these excitations can be observed experimentally.
Note also that the $K=1$ skyrmion and anti-skyrmion states are 
perfect analogs of the interband charged exciton states, except 
that $e$ and $h$ are replaced by QE$_{\rm R}$ and QH.
In analogy to $\nu=1$, the values of $E_S(1)-E_0$ measure the 
binding energies of such (fractionally) charged CF spin-excitons, 
$X_{\rm CF}^-=2{\rm QE}_{\rm R}+{\rm QH}$ and $X_{\rm CF}^+=
2{\rm QH}+{\rm QE}_{\rm R}$.

It might seem surprising that the spin excitations at $\nu=1$ and 
${1\over3}$ are similar despite different interactions between 
electrons and CF's.
However, of all three types of QP's at $\nu={1\over3}$, 
only QH and QE$_{\rm R}$ are involved in the low-energy spin 
excitations $W_K$ and $S_K$, and the QH--QH, QH--QE$_{\rm R}$, 
and QE$_{\rm R}$--QE$_{\rm R}$ pseudopotentials describing their 
interactions,\cite{hierarchy,qer} are all quite similar to the 
$e$--$e$ and $e$--$h$ pseudopotentials in the lowest electron LL.
On the other hand, the QE's, whose interaction (at short range) 
with one another and with other QP's is very different,\cite{hierarchy} 
do not participate in $W_K$ and $S_K$ excitations.

\section{Integral quantum Hall regime (excited Landau levels)}

Another system with an identical structure of the single-particle 
Hilbert space but with different interactions is a nearly completely 
filled excited ($n>0$) LL, experimentally realized in the 2DEG at 
$\nu\approx2n+1$.
Note that if skyrmions would indeed occur at $\nu=3$, 5, etc., 
they should be observed even more easily than at $\nu=1$ because 
of the weaker magnetic field $B$ (at the same 2DEG density), 
and thus smaller $\eta=E_{\rm Z}/E_{\rm C}\propto\sqrt{B}$.

\begin{figure}[t]
\epsfxsize=3.40in
\epsffile{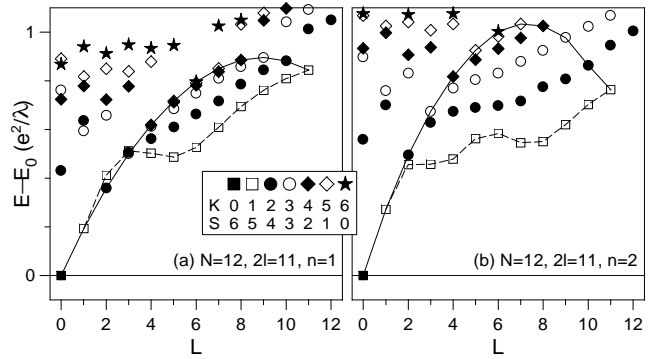}
\caption{
   Same as Fig.~\ref{fig01}(a) but for electrons confined to
   the first (a) and second (b) excited LL.}
\label{fig05}
\end{figure}

\begin{figure}[t]
\epsfxsize=3.40in
\epsffile{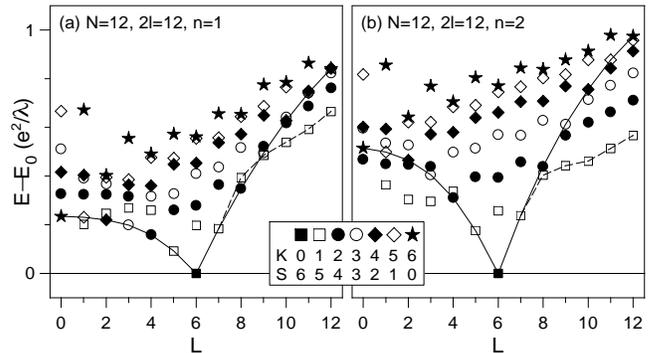}
\caption{
   Same as Fig.~\ref{fig02}(a) but for electrons confined to
   the first (a) and second (b) excited LL.}
\label{fig06}
\end{figure}

In this section we shall discuss the results for an ideal 2D system 
with zero layer width, $w=0$.
The energy spectra analogous to those in Figs.~\ref{fig01}(a) 
and \ref{fig02}(a) but calculated for $n=1$ and 2 are shown in 
Figs.~\ref{fig05} and \ref{fig06}.
The same the number of electrons $N=12$ (in the $n$th LL; 
the lower LL's are completely filled) and the angular momenta 
$l=11$ and 12 have been chosen, yielding the monopole strength 
$2Q=2(l-n)$.
Clearly, none of the above-discussed features of the $\nu=1$ or 
$n={1\over3}$ ($\nu^*=1$) spectra are present at $\nu=3$ or 5.

Let us begin with Fig.~\ref{fig05} for $N=g$.
A single SW (dashed lines; for dispersion see 
Ref.~\onlinecite{Kallin84}) is generally the lowest-energy spin 
excitation (at any $L$), and the $W_K$ bands (identified by 
comparison of the pair correlation functions) have higher energy 
and are no longer linear.
The sublinear and nearly parabolic $E_W\approx E_0+u\zeta
+y\zeta^2$ can be interpreted as an attraction between the $L=1$ 
SW's and, based on data for $N\le14$, we find $u=2.41\,E_{\rm C}$ 
and $y=-1.62\,E_{\rm C}$ for $n=1$, and $u=3.45\,E_{\rm C}$ and 
$y=-2.88\,E_{\rm C}$ for $n=2$.
In a finite-size $\nu=3$ or 5 droplet, as a result of this attraction, 
the spin-singlet condensate of $K={1\over2}N$ SW's (marked with arrows 
in Fig.~\ref{fig01}) is an excited state at any strength of confinement 
($\Omega$), and the edge-reconstruction of the $\nu=3$ or 5 ferromagnetic 
GS ($C_0$) occurs directly to the next compact-droplet state, $C_1$.
This different behavior might be probed in a transport experiment
by sending a reversed-spin electron over a quantum dot containing 
a compact droplet.
It seems that a reversed-spin carrier would induce and bind SW's 
when sent through a $\nu=1$ or ${1\over3}$ droplet, and travel 
ballistically through a $\nu=3$ or 5 droplet.

The lack of response to an addition of a reversed-spin electron
(from the edge into the inside of the droplet) must mean unbinding 
of skyrmions at $\nu=3$ or 5.
Indeed, the $S_K$ states in the spectra for $N+1=g$ in Fig.~\ref{fig06}
all have $E>E_0$.
This means no skyrmions in the excited LL's at any value of $E_{\rm Z}$.
In contrast to the situation near $\nu=1$ or ${1\over3}$, the GS both 
precisely at $\nu=3$ or 5 and in the vicinity of these values remains 
maximally spin-polarized even in the absence of Zeeman splitting.
In the $e$--$h$ picture, the result is that no bound charged-exciton
states $X_K^-$ occur in excited LL's (in the absence of inter-LL mixing
and finite well width effects).

\section{Effects of finite layer width}

\begin{figure}[t]
\epsfxsize=3.40in
\epsffile{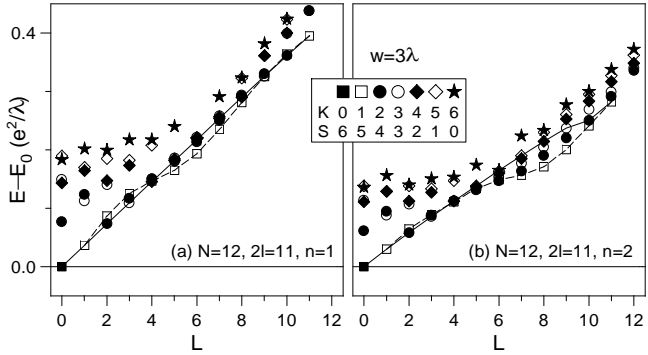}
\caption{
   Same as Fig.~\ref{fig05} but for a finite width $w=3\lambda$ of 
   a quasi-2D electron layer.}
\label{fig07}
\end{figure}

\begin{figure}[t]
\epsfxsize=3.40in
\epsffile{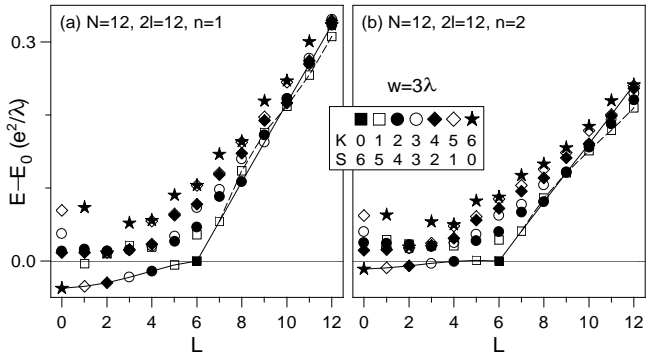}
\caption{
   Same as Fig.~\ref{fig06} but for a finite width $w=3\lambda$ of 
   a quasi-2D electron layer.}
\label{fig08}
\end{figure}

As was first predicted by Cooper\cite{Cooper97} and later confirmed 
experimentally by Song {\sl et al.},\cite{Song99} skyrmions become 
the lowest-energy charged excitations in the excited LL's as well, 
if only the layer width $w$ is sufficiently large.
We have calculated the spin-excitation spectra analogous to those 
of Figs.~\ref{fig05} and \ref{fig06} but for $w=3\lambda$, and show
them in Figs.~\ref{fig07} and \ref{fig08}.
For the exact fillings of the $n=1$ and 2 LL's ($\nu=3$ and 5;
Fig.~\ref{fig07}), the $E_W(K)$ energy bands which were strongly 
sublinear for $w=0$ now become nearly linear (similar to the 
lowest LL; see Fig.~\ref{fig01}).
This indicates vanishing of the SW--SW attraction, and reoccurrence
of the Bose condensate of ordered $L=1$ SW's.
For an additional hole in the $n=1$ and 2 LL's ($\nu=3^-$ and $5^-$; 
Fig.~\ref{fig08}), the skyrmion energy bands which had $E_S(K)>E_0$ 
for $w=0$ now have $E_S(K)<E_0$ (again, similar to the lowest LL; 
see Fig.~\ref{fig02}).
This indicates stability of skyrmions in excited LL's in a wide 
quasi-2D layer (at sufficiently small $E_{\rm Z}$).
Also in Fig.~\ref{fig08}, the $L=N-S$ bands of states (containing 
$K$ SW's created coherently in the presence of a hole) become now 
nearly linear, in contrast to the behavior in Fig.~\ref{fig06} but 
similarly to Fig.~\ref{fig02}.

\begin{figure}[t]
\epsfxsize=3.40in
\epsffile{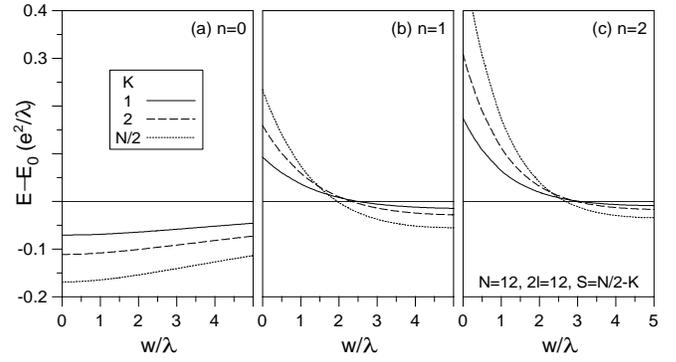}
\caption{
   The energy $E$ of skyrmions with $K=1$, 2, and ${1\over2}N$ 
   as a function of the layer width $w$, calculated 
   on Haldane sphere for $N=12$ electrons in the lowest (a), 
   first excited (b), and second excited (c) LL.
   $\lambda$ is the magnetic length.}
\label{fig09}
\end{figure}

In Fig.~\ref{fig09} we compare the energies of skyrmions with $K=1$, 
2, and ${1\over2}N$ (the latter state has $S=0$ and would correspond 
to an infinite-size skyrmion in the $N=\infty$ limit) plotted as 
a function of the layer width $w$.
Clearly, the skyrmion energy is more sensitive to $w$ in the 
excited LL's.
The ``binding energies'' $E_S(K)-E_0$ that were all positive 
for $w=0$ in Fig.~\ref{fig06} change sign at $w/\lambda=2$ to 3, 
depending on $K$ and $n$.
Note that our critical values of $w$ are quite higher than those 
predicted by Cooper (for example, for the $n=1$ LL, his critical 
parameter $a=0.09\lambda$ for the gaussian density profile, 
$\varrho(z)\propto\exp(-z^2/2a^2)$ corresponds to $w\approx
0.5\lambda$ for our $\varrho(z)\propto\cos^2(z\pi/w)$).
This discrepancy indicates slow convergence of the energy of an 
infinite ($S=0$) skyrmion with the electron number $N$.
However, our critical values are certainly more appropriate for 
small skyrmions which are observed experimentally.
Let us compare these values with a pair of experiments in which 
the skyrmions were and were not observed at $\nu=3$.
Taking parameters after Song {\sl et al.},\cite{Song99} who observed 
skyrmions with $K\le2$ ($B=2.15$~T and well width of 30~nm yielding 
$w=33.5$~nm) gives $w/\lambda=1.9$, just above our critical value 
(see Fig.~\ref{fig11} for data for $N\rightarrow\infty$).
On the other hand, taking the parameters after Schmeller 
{\sl et al.},\cite{Schmeller95} who did not observe skyrmions 
($B=2.3$~T and well widths of 14 and 20~nm yielding $w=17.5$ and 
23.5~nm) gives $w/\lambda=1.03$ and 1.40 for two different samples, 
both significantly below our critical value and above that given 
by Cooper.

\begin{figure}[t]
\epsfxsize=3.40in
\epsffile{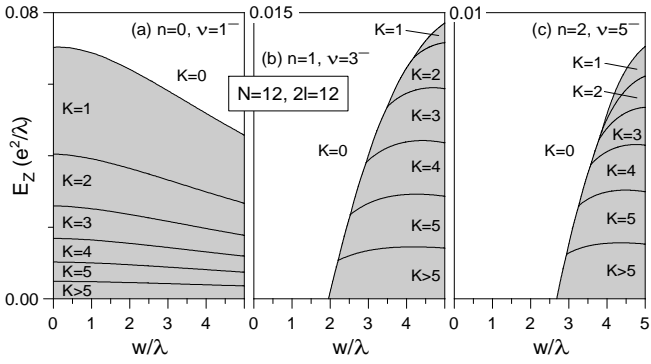}
\caption{
   Phase diagrams for the occurrence of skyrmions with $K$ 
   reversed spins in a quasi-2D electron gas of finite width $w$,
   calculated in the system of $N=12$ electrons in the lowest (a), 
   first excited (b), and second excited (c) LL of degeneracy 
   $g=2l+1=12$.
   $E_{\rm Z}$ is the Zeeman energy and $\lambda$ is the magnetic 
   length.
   The numbers in the top-left corners of frames (b) and (c) give
   the upper bounds of their vertical axes (the lower bound are
   zero in all frames).}
\label{fig10}
\end{figure}

\begin{figure}[t]
\epsfxsize=3.40in
\epsffile{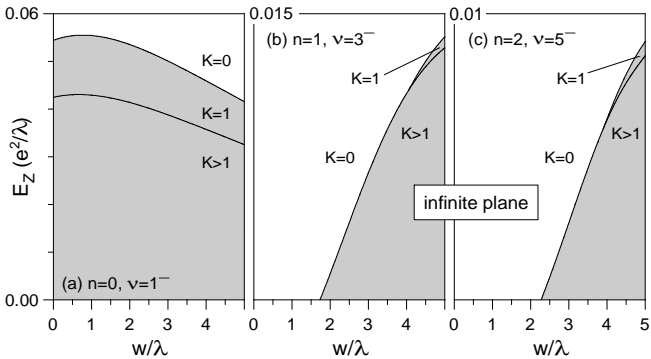}
\caption{
   Same as Fig.~\ref{fig10} but for an infinite planar system
   (critical values of Zeeman energy $E_{\rm Z}$ for each layer 
   width $w$ were obtained from extrapolation of data for electron
   numbers $N\le50$ to the limit of $N\rightarrow\infty$).}
\label{fig11}
\end{figure}

Using the data of Fig.~\ref{fig09} one can calculate a phase diagram
for the occurrence of skyrmions with a given number of reversed 
spins $K$ as a function of the layer width $w$ and the Zeeman energy 
$E_{\rm Z}$.
Such a diagram is presented in Fig.~\ref{fig10} for the integral 
filling of the lowest and first two excited LL's ($\nu=1$, 3, and 
5, respectively).
To have a more reliable estimate of the critical $w/\lambda$ and 
$\eta$ for the occurrence of skyrmions with any $K\le1$, we have
recalculated the curves for $K=1$ and 2 for much larger $N$ (up 
to 50) and then, thanks to their regular dependence on $N$, were
able to extrapolate them to the $N\rightarrow\infty$ limit.
The resulting phase diagram, shown in Fig.~\ref{fig11}, describes 
an infinite planar system, and it is consistent with the skyrmion 
energies reported by Palacios {\sl et al.}\cite{Palacios96} for
$n=0$ and $w=0$.
\begin{figure}[t]
\epsfxsize=3.40in
\epsffile{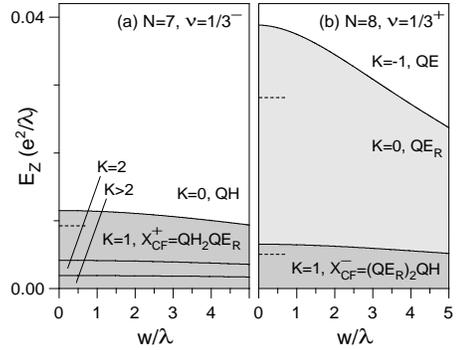}
\caption{
   Same as Fig.~\ref{fig10} but for the fractional quantum Hall
   states near $\nu={1\over3}$.
   Horizontal dashes lines mark the critical values of $E_{\rm Z}$ 
   at $w=0$ obtained from extrapolation of the finite-size data for 
   $N\le9$ to the $N\rightarrow\infty$ limit.}
\label{fig12}
\end{figure}
Remarkably, the critical value of $E_{\rm Z}$ for the lowest LL 
is quite insensitive to $w$ over a wide range of layer widths.
This is in contrast to the situation in the excited LL's, for
which the phase diagrams in Fig.~\ref{fig11}(bc) show a similar 
fast increase of the critical $E_{\rm Z}$ with increasing $w$.
The critical layer widths in the limit of vanishing Zeeman energy
are $w/\lambda=1.8$ and 2.3 for $n=1$ and 2, respectively.
In view of a recent study\cite{Melik99} that showed that the LL 
mixing only weakly affects the skyrmion energies in the layers of 
nonzero width, we expect our phase diagrams in Fig.~\ref{fig11} 
to be quite adequate for realistic experimental systems.

Finally, in Fig.~\ref{fig12} we present an analogous phase 
diagram for the $\nu={1\over3}$ fractional quantum Hall state.
Due to the broken QE$_{\rm R}$--QH symmetry, the diagrams for 
skyrmions (at $\nu={1\over3}^+$) and anti-skyrmions (at $\nu=
{1\over3}^-$) are different, and they are both shown.
The solid lines and shaded areas give the result for small systems:
$N=7$ and $2l=19$ in frame (a), and $N=8$ and $2l=20$ in frame (b).
The dashed lines give the critical values of $E_{\rm Z}$ at $w=0$ 
for the $X_{\rm CF}^+$, QE$_{\rm R}$, and $X_{\rm CF}^-$ states,
obtained from extrapolation of data for $N\le9$ to $N\rightarrow
\infty$.

\section{General criterion for skyrmions in half-filled 
         spin-degenerate shell}

\begin{figure}[t]
\epsfxsize=3.40in
\epsffile{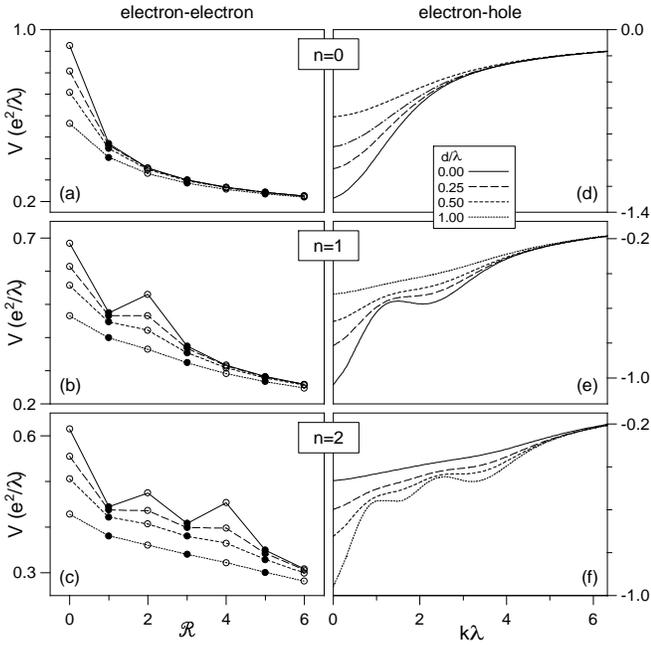}
\caption{
   Pseudopotentials $V$ of the $e$--$e$ (abc) and $e$--$h$ 
   (def) interaction in the $n=0$ (ad), 1 (be), and 2 (cf) LL's, 
   calculated using the interaction potential 
   $V_d(r)=e^2/\sqrt{r^2+d^2}$ for $d/\lambda=0$, ${1\over4}$, 
   ${1\over2}$, and 1, and describing a quasi-2D layer of width 
   $w=5d$.
   ${\cal R}$ is relative $e$--$e$ angular momentum (data shown only
   for ${\cal R}\le6$), open and closed circles mark singlet and 
   triplet $e$--$e$ states, respectively, $k$ is the $e$--$h$ wave 
   vector, and $\lambda$ is the magnetic length.}
\label{fig13}
\end{figure}

Since 
(i) the finite width $w$ enters the Hamiltonian (\ref{eqham}) only 
through the pseudopotential $V({\cal R})$, 
(ii) only a few leading parameters $V(0)$, $V(1)$, $V(2)$, \dots\ 
that correspond to a short average $e$--$e$ distance 
$\sqrt{\left<r^2\right>}$ significantly depend on $w$, and
(iii) for $w=0$, the opposite behavior of $E_S(K)-E_0$ for the lowest 
and excited LL's results precisely from the enhanced value of $V(2)$ 
for $n\ge1$, it is enough to study the dependence of the short-range 
part of $V({\cal R})$ on $w$ to understand the reoccurrence of
skyrmions for $n\ge1$ at $w\sim2\lambda$.
This dependence is illustrated in Fig.~\ref{fig13}(abc).
Clearly, increasing $d$ (i.e., $w=5d$) suppresses more strongly the 
pseudopotential parameters at the even values of ${\cal R}$ (open 
circles) corresponding to zero pair spin, specially the highly 
repulsive ones at ${\cal R}=0$ and 2 for $n=1$ and ${\cal R}=0$, 2,
and 4 for $n=2$.
While in any LL the strong suppression of $V(0)$ will eventually 
(at very large $w$) destroy skyrmions (all having no pairs with
${\cal R}=0$), there is a wide range of $w$ in which skyrmions 
become stable also for $n>0$.

By comparing the values of $w$ and $d$ in Figs.~\ref{fig09}(abc) 
and \ref{fig13}(abc), we find the following general conditions 
for the occurrence of skyrmions in a system of interacting 
spin-${1\over2}$ fermions half-filling a spin-degenerate shell:
(i) $V(0)$ must be sufficiently large to cause decoupling of the 
many-body states without the ${\cal R}=0$ pairs (skyrmions) from 
all other many-body states, and
(ii) $V({\cal R})$ must decrease with increasing ${\cal R}\ge1$.

Note that the latter condition (ii) is not immediately applicable 
to shells with broken particle($\uparrow$)--hole($\downarrow$) 
symmetry.
Examples of such systems include electrons at $\nu={1\over3}$ 
(i.e., CF's at $\nu^*=1$) but also the case when the pair of 
spin-degenerate LL's has different orbital indices $n$ (which 
can be realized by making the Zeeman gap $E_Z$ equal to the 
cyclotron gap $\hbar\omega_c$ in a magnetic material).
In these systems, the interaction Hamiltonian (\ref{eqham}) 
is determined by a pair of (different) particle--particle and 
hole--hole pseudopotentials $V_{\uparrow\uparrow}({\cal R})$ and 
$V_{\downarrow\downarrow}({\cal R})$, where ${\cal R}$ is an odd 
number (as required for two identical fermions), and a 
particle--hole continuous dispersion $V_{\uparrow\downarrow}(k)$, 
where $k$ is the pair wave vector (on a sphere, $kR=L$).
To rephrase condition (ii) in terms of $V_{\uparrow\downarrow}(k)$, 
we notice in Fig.~\ref{fig13} that the suppression of the maxima at 
${\cal R}=2$ (for $n=1$) and ${\cal R}=2$ and 4 (for $n=2$) coincides 
with the disappearance of the corresponding roton minima in the 
spin-wave dispersion $V_{\rm eh}(k)$, at $k\lambda\approx2.1$ 
(for $n=1$) and $k\lambda\approx1.5$ and 3.2 (for $n=2$).

A continuous evolution of the skyrmion energy spectrum $E_S(K)$
from the positive values (as for $n\ge1$ and small $w$) to the 
negative values (as for $n=0$ or $n\ge1$ and large $w$) can be
most easily understood by studying a simple model pseudopotential 
$U_x({\cal R})$ defined as: $U_x(0)=\infty$, $U_x(1)=1$, 
$U_x(2)=x$, and $U_x({\cal R})=0$ for ${\cal R}>2$.
This choice of $U_x$ guarantees that skyrmions are its only 
finite-energy eigenstates, and their energy spectrum $E_S(K)$
depends on one free parameter $x$.

\begin{figure}[t]
\epsfxsize=3.40in
\epsffile{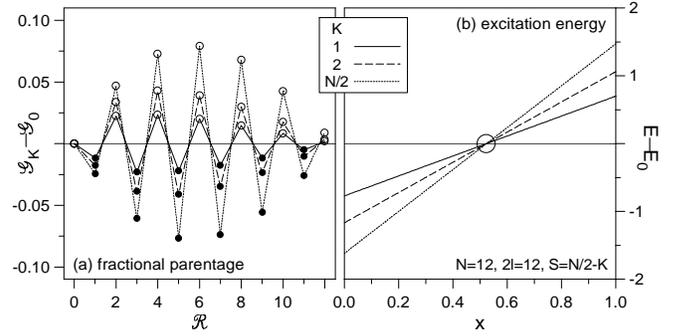}
\caption{
   (a) 
   Pair-correlation functions -- fractional grandparentage 
   ${\cal G}_K$ as a function of relative pair angular momentum 
   ${\cal R}$ -- for skyrmions with $K=1$ and ${1\over2}N$, 
   calculated for $N=12$ electrons on Haldane sphere.
   Open and closed circles mark singlet and triplet pair states, 
   respectively.
   (b)
   Energy $E$ of the skyrmions with $K=1$, 2, and ${1\over2}N$ as a 
   function of parameter $x$ of the model $e$--$e$ interaction $U_x$.}
\label{fig14}
\end{figure}

The essential information about the skyrmion wave functions is 
contained in the fractional grandparentage coefficients
\cite{acta,philmag} ${\cal G}$.
The dependence of ${\cal G}$ on ${\cal R}$ is a pair correlation 
function defined as a fraction of the $e$--$e$ pairs with the 
relative pair angular momentum ${\cal R}$.
For $K=0$, the many-electron system is completely spin-polarized, 
so that every electron pair has spin $S=1$, and thus ${\cal G}$ 
vanishes for all even values of ${\cal R}$.
When $K$ is increased, so does the total fraction of $S=0$ 
pairs, ${\cal G}_K(0)+{\cal G}_K(2)+\dots$, which happens at 
the cost of a decreasing number of $S=1$ pairs, ${\cal G}_K(1)+
{\cal G}_K(3)+\dots$.
The grandparentage coefficients measured from the ``reference'' 
value ${\cal G}_0$ for $K=0$ are plotted in Fig.~\ref{fig14}(a) 
as a function of ${\cal R}$ for $K=1$ and ${1\over2}N$. 
It turns out that $\Delta{\cal G}_K={\cal G}_K-{\cal G}_0$ is 
a regular function of ${\cal R}$ and, for example, 
$\Delta{\cal G}_K(1)=-\alpha{\cal G}_K(2)$, where $\alpha^{-1}=
2-(N-1)^{-1}$.
This allows using a general expression for the interaction energy,
\cite{acta,philmag} 
\begin{equation}
   E={1\over2}N(N-1)\sum_{\cal R}{\cal G}({\cal R})V({\cal R}), 
\end{equation}
to write the skyrmion energy for $V=U_x$ as 
\begin{equation}
   E_S(K)-E_0=(x-\alpha){\cal G}_K(2).
\end{equation}
As shown in Fig.~\ref{fig14}(b), $E_S(K)-E_0$ changes sign simultaneously 
for all $K$ (the spin-polarized $K=0$ QP state becomes unstable toward 
formation of a skyrmion) when $x=\alpha$, that is when $U_x(2)$ drops 
below $\alpha U_x(1)$.
Since $\alpha\rightarrow{1\over2}$ for $N\rightarrow\infty$, this
means that skyrmions in an infinite (planar) system will have lower 
energy than a QP state with $K=0$ when $U_x$ becomes superlinear 
(i.e., superharmonic\cite{jphys,acta,philmag}) between ${\cal R}=1$ 
and 3.

\section{Conclusion}

We have presented the results of detailed numerical studies of 
the spin excitations of various ferromagnetic GS's of a 2DEG 
confined in a quantum well of finite width $w$, in the integral 
and fractional quantum Hall regime.
Similar low-energy spectra in the vicinity of $\nu=1$ and 
${1\over3}$ ($\nu^*=1$ in the CF picture) have been found, and 
both contain the following two types of low-energy excitations: 
spin waves\cite{Kallin84}, and skyrmions and anti-skyrmions
\cite{Sondhi93,MacDonald98}.
The phase diagrams for the occurrence of skyrmions with different
numbers of reversed spins $K$ as a function of the well width $w$ 
and the Zeeman energy $E_{\rm Z}$ have been determined at $\nu=1$,
3, and 5.

The interactions between the (neutral) spin waves and (charged) 
skyrmions have also been studied.
We found that the spin waves each carrying angular momentum $L=1$ 
condense into an ordered (with parallel angular momenta), correlated, 
and noninteracting state.
The interaction energy $E_W$ of this condensate is a linear function 
of (continuous) spin polarization $\zeta$ which, in the absence of 
the Zeeman energy, gives rise to a gapless and continuous density 
of states.
This is in contrast to a discrete spectrum of particle-like skyrmion 
excitations, whose energy $E_S$ is a function of the (integral) 
reversed-spin number $K$.
The short-range repulsion between charged skyrmions is predicted 
to cause Laughlin correlations, that is the tendency to avoid skyrmion 
pair eigenstates with the smallest relative angular momenta ${\cal R}$.
This causes effective spatial isolation of skyrmions from one another, 
and the absence of high-energy skyrmion--skyrmion collisions.

The major differences between the $\nu=1$ and ${1\over3}$ spectra
are the reduced energy scale and a broken skyrmion--anti-skyrmion
symmetry in the latter system (broken particle--hole symmetry
in the lowest CF LL).
A number of phenomena associated with the particular form of spin 
excitations at $\nu\approx1$ (rapid depolarization at $\nu=1^\pm$, 
nonlinear transport through a finite-size droplet, sensitivity of 
the spin coupling to magnetic particles to $\nu$, etc.) are also 
expected at $\nu\approx{1\over3}$.
The smallest skyrmion and anti-skyrmion states at $\nu={1\over3}$ 
are equivalent to composite fermion charged excitons $X_{\rm CF}^\pm$.

A qualitatively different behavior is observed in the excited LL's.
In agreement with earlier theories,\cite{Jain94,Wu95} we find that 
skyrmions and anti-skyrmions are unstable at $\nu=3$ or 5 even at 
$E_{\rm Z}=0$, which (in contrast to $\nu=1$ or ${1\over3}$) results 
in the stability of the ferromagnetic order at all nearby values of 
$\nu$, and the single-particle character of elementary excitations.
In the $e$--$h$ picture, this means unbinding of charged excitons
in the excited LL's.
Also in contrast to $\nu=1$ or ${1\over3}$, the $L=1$ spin waves 
attract one another rather than decouple, which for example 
results in direct confinement-induced transitions between the 
consecutive ``compact'' states of finite $\nu=3$ or 5 quantum Hall 
droplets, skipping the correlated, depolarized states with 
intermediate density.

This different behavior in the excited LL's is suppressed when
the width $w$ of a quasi-2D layer exceeds about two magnetic lengths.
This critical value obtained from a finite-size calculation seems 
to agree better with the experiments\cite{Schmeller95,Song99} than 
an earlier estimate.\cite{Cooper97}
The reoccurrence of skyrmions in excited LL's in wider quantum
wells is explained by studying the involved particle--particle 
and particle--hole interaction pseudopotentials and the electron
correlations in the skyrmion eigenstates.

A general criterion is found that allows prediction of the 
presence or absence of skyrmions in a system of interacting 
spin-${1\over2}$ fermions in a half-filled spin-degenerate shell.
The criterion describes correctly all calculated spin excitation 
spectra at $\nu={1\over3}$, 1, 3, and 5, and at arbitrary layer 
widths $w$, density profile $\varrho(z)$, etc.

\section*{Acknowledgment}
The authors acknowledge partial support by the Materials Research 
Program of Basic Energy Sciences, US Department of Energy.
JJQ thanks National Magnetic Field Laboratory, Tallahassee, 
and University of New South Wales, Sydney, for hospitality.
AW  acknowledges helpful discussions with P. Hawrylak, L. Jacak, 
A. H. MacDonald, J. J. Palacios, M. Potemski, and I. Szlufarska, 
and support from grant 2P03B11118 of the Polish State Committee 
for Scientific Research (KBN).


\begin{references}

\bibitem{Prange87}
{\sl The Quantum Hall Effect},
   edited by R. E. Prange and S. M. Girvin 
   (Springer-Verlag, New York, 1987).

\bibitem{Chakraborty95}
T. Chakraborty and P. Pietil\"ainen,
   {\sl The Quantum Hall Effects}, 
   (Springer-Verlag, New York, 1995).

\bibitem{DasSarma97}
{\sl Perspectives in Quantum Hall Effects},
   edited by S. Das Sarma and A. Pinczuk,
   (John Wiley \& Sons, New York, 1997).

\bibitem{Laughlin83}
R. B. Laughlin,
   Phys. Rev. Lett. {\bf50}, 1395 (1983).

\bibitem{Haldane87}
F. D. M. Haldane, 
  in Ref.~\onlinecite{Prange87}, chap.~8, p.~303.

\bibitem{jphys}
J. J. Quinn and A. W\'ojs,
   J. Phys.: Cond. Mat. {\bf12}, R265 (2000).

\bibitem{Tsui82}
D. C. Tsui, H. L. Stormer, and A. C. Gossard, 
   Phys. Rev. Lett. {\bf48}, 1559 (1982).

\bibitem{Barret95}
S. E. Barrett, G. Dabbagh, L. N. Pfeiffer, K. W. West, and R. Tycko,
   Phys. Rev. Lett. {\bf74}, 5112 (1995).

\bibitem{Tycko95}
R. Tycko, S. E. Barrett, G. Dabbagh, L. N. Pfeiffer, and K. W. West, 
   Science {\bf268}, 1460 (1995).

\bibitem{Maude96}
D. K. Maude, M. Potemski, J. C. Portal, M. Henini, L. Eaves, 
G. Hill, and M. A. Pate, 
   Phys. Rev. Lett. {\bf77}, 4604 (1996).

\bibitem{Aifer96}
E. H. Aifer, B. B. Goldberg, and D. A. Broido,
   Phys. Rev. Lett. {\bf76}, 680 (1996).

\bibitem{Manfra96}
M. J. Manfra, E. H. Aifer, B. B. Goldberg, D. A. Broido, 
L. Pfeiffer, and K. West,
   Phys. Rev. B {\bf54}, R17327 (1996).

\bibitem{Shukla00}
S. P. Shukla, M. Shayegan, S. R. Parihar, S. A. Lyon,
N. R. Cooper, and A. A. Kiselev,
   Phys. Rev. B {\bf61}, 4469 (2000).

\bibitem{Khandelwal01}
P. Khandelwal, A. E. Dementyev, N. N. Kuzma, S. E. Barrett, 
L. N. Pfeiffer, and K. W. West,
   Phys. Rev. Lett. {\bf86}, 5353 (2001).

\bibitem{Skyrme61}
T. Skyrme,
   Proc. R. Soc. London A {\bf262}, 237 (1961). 

\bibitem{Rajaraman82}
R. Ramajaran,
   {\sl Solitons and Instantons},
   (North-Holland, Amsterdam, 1982).

\bibitem{Lee90}
D. H. Lee and C. L. Kane, 
   Phys. Rev. Lett. {\bf64}, 1313 (1990). 

\bibitem{Haldane83}
F. D. M. Haldane,
   Phys. Rev. Lett. {\bf51}, 605 (1983).

\bibitem{Jain89}
J. K. Jain,
   Phys. Rev. Lett. {\bf63}, 199 (1989).

\bibitem{Lopez91}
A. Lopez and E. Fradkin,
   Phys. Rev. B {\bf44}, 5246 (1991).

\bibitem{Halperin93}
B. I. Halperin, P. A. Lee, and N. Read,
   Phys. Rev. B {\bf47}, 7312 (1993).

\bibitem{Sitko97}
P. Sitko, K. S. Yi, and J. J. Quinn,
   Phys. Rev. B {\bf56}, 12417 (1997).
  
\bibitem{Klitzing80}
K. von Klitzing, G. Dorda, and M. Pepper,
   Phys. Rev. Lett. {\bf45}, 494 (1980).

\bibitem{acta}
A. W\'ojs and J. J. Quinn, 
   Acta Phys. Pol. A {\bf96}, 403 (1999).

\bibitem{philmag}
A. W\'ojs and J. J. Quinn,
   Philos. Mag. B {\bf80}, 1405 (2000).

\bibitem{fivehalf}
A. W\'ojs,
   Phys. Rev. B {\bf63}, 125312 (2001).

\bibitem{hierarchy}
A. W\'ojs and J. J. Quinn,
   Phys. Rev. B {\bf61}, 2846 (2000).

\bibitem{He90}
S. He, F. C. Zhang, X. C. Xie, and S. Das Sarma, 
   Phys. Rev. B {\bf42}, R11376 (1990).

\bibitem{Sondhi93}
S. L. Sondhi, A. Karlhede, S. A. Kivelson, and E. H. Rezayi,
   Phys. Rev. B {\bf47}, 16419 (1993).

\bibitem{Fertig94}
H. A. Fertig, L. Brey, R. C\^ot\'e, and A. H. MacDonald,
   Phys. Rev. B {\bf50}, 11018 (1994).

\bibitem{Brey95}
L. Brey, H. A. Fertig, R. C\^ot\'e, and A. H. MacDonald,
   Phys. Rev. Lett. {\bf75}, 2562 (1995).

\bibitem{Moon95}
K. Moon, H. Mori, K. Yang, S. M. Girvin, A. H. MacDonald,
L. Zheng, D. Yoshioka, and S. C. Zhang,
   Phys. Rev. B {\bf51}, 5138 (1995).

\bibitem{MacDonald96}
A. H. MacDonald, H. A. Fertig, and L. Brey, 
   Phys. Rev. Lett. {\bf76}, 2153 (1996).

\bibitem{Xie96}
X. C. Xie and S. He, 
   Phys. Rev. B {\bf53}, 1046 (1996).

\bibitem{Cote96}
R. C\^ot\'e and A. H. MacDonald,
   Phys. Rev. B {\bf53}, 10019 (1996).

\bibitem{Cote97}
R. C\^ot\'e, A. H. MacDonald, L. Brey, H. A. Fertig, 
S. M. Girvin, and H. T. C. Stoof,
   Phys. Rev. Lett. {\bf78}, 4825 (1997).

\bibitem{Abolfath97}
M. Abolfath, J. J. Palacios, H. A. Fertig, S. M. Girvin, 
and A. H. MacDonald,
   Phys. Rev. B {\bf56}, 6795 (1997).

\bibitem{Fertig97}
H. A. Fertig, L. Brey, R. C\^ot\'e, A. H. MacDonald, A. Karlhede, 
and S. L. Sondhi,
   Phys. Rev. B {\bf55}, 10671 (1997).

\bibitem{Oaknin98}
J. H. Oaknin, B. Paredes, and C. Tejedor, 
   Phys. Rev. B {\bf58}, 13028 (1998).

\bibitem{Melik99}
V. Melik-Alaverdian, N. E. Bonesteel, and G. Ortiz,
   Phys. Rev. B {\bf60}, R8501 (1999).

\bibitem{Palacios94}
J. J. Palacios, L. Martin-Moreno, G. Chiappe, E. Louis, and C. Tejedor,
   Phys. Rev. B {\bf50}, 5760 (1994).

\bibitem{Oaknin96}
J. H. Oaknin, L. Martin-Moreno, and C. Tejedor,
   Phys. Rev. B {\bf54}, 16850 (1996).

\bibitem{droplet}
A. W\'ojs and P. Hawrylak,
   Phys. Rev. B {\bf56}, 13227 (1997).

\bibitem{MacDonald90}
A. H. MacDonald and E. H. Rezayi,
   Phys. Rev. B {\bf42}, 3224 (1990).

\bibitem{x-td}
A. W\'ojs and P. Hawrylak,
   Phys. Rev. B {\bf51}, 10880 (1995).

\bibitem{Palacios96}
J. J. Palacios, D. Yoshioka, and A. H. MacDonald,
   Phys. Rev. B {\bf54}, 2296 (1996).

\bibitem{Kamilla96}
R. K. Kamilla, X. G. Wu, and J. K. Jain,
   Solid State Commun. {\bf99}, 289 (1996).

\bibitem{Leadley97}
D. R. Leadley, R. J. Nicholas, D. K. Maude, A. N. Utjuzh, 
J. C. Portal, J. J. Harris, and C. T. Foxon,
   Phys. Rev. Lett. {\bf79}, 4246 (1997).

\bibitem{Khandelwal98}
P. Khandelwal, N. N. Kuzma, S. E. Barrett, L. N. Pfeiffer, 
and K. W. West,
   Phys. Rev. Lett. {\bf81}, 673 (1998).

\bibitem{MacDonald98}
A. H. MacDonald and J. J. Palacios,
   Phys. Rev. B {\bf58}, R10171 (1998).

\bibitem{Jain94}
J. K. Jain and X. G. Wu,
   Phys. Rev. B {\bf49}, 5085 (1994).

\bibitem{Wu95}
X. G. Wu and S. L. Sondhi,
   Phys. Rev. B {\bf51}, 14725 (1995).

\bibitem{Schmeller95}
A. Schmeller, J. P. Eisenstein, L. N. Pfeiffer, and K. W. West, 
   Phys. Rev. Lett. {\bf75}, 4290 (1995).

\bibitem{Cooper97}
N. R. Cooper,
   Phys. Rev. B {\bf55}, R1934 (1997).

\bibitem{Song99}
Y. Q. Song, B. M. Goodson, K. Maranowski, and A. C. Gossard,
   Phys. Rev. Lett. {\bf82}, 2768 (1999).

\bibitem{x-fqhe}
A. W\'ojs, P. Hawrylak, and J. J. Quinn,
   Phys. Rev. B {\bf60}, 11661 (1999).

\bibitem{x-cf}
A. W\'ojs, I. Szlufarska, K. S. Yi, and J. J. Quinn,
   Phys. Rev. B {\bf60}, R11273 (1999).

\bibitem{x-tb}
A. W\'ojs, J. J. Quinn, and P. Hawrylak,
   Phys. Rev. B {\bf62}, 4630 (2000).

\bibitem{qer}
I. Szlufarska, A. W\'ojs, and J. J. Quinn,
   Phys. Rev. B {\bf64} (to appear October 15, 2001).

\bibitem{Chakraborty86}
T. Chakraborty, P. Pietil\"ainen, and F. C. Zhang, 
   Phys. Rev. Lett. {\bf57}, 130 (1986).

\bibitem{Rezayi87}
E. H. Rezayi, 
   Phys. Rev. B {\bf36}, 5454 (1987);
   {\sl ibid.} {\bf43}, 5944 (1991). 

\bibitem{Wu76}
T. T. Wu and C. N. Yang,
   Nucl. Phys. B {\bf107}, 365 (1976).

\bibitem{Fano86}
G. Fano, F. Ortolani, and E. Colombo,
   Phys. Rev. B {\bf34}, 2670 (1986).

\bibitem{Avron78}
J. E. Avron, I. W. Herbst, and B. Simon,
   Ann. Phys. {\bf114}, 431 (1978).
  
\bibitem{geometry}
A. W\'ojs and J. J. Quinn, 
   Physica E {\bf3}, 181 (1998).

\bibitem{Kallin84}
C. Kallin and B. I. Halperin,
   Phys. Rev. B {\bf30}, 5655 (1984).

\bibitem{Dzyubenko83}
A. B. Dzyubenko and Yu. E. Lozovik,
   Fiz. Tverd. Tela {\bf25}, 1519 (1983)
   [Sov. Phys. Solid State {\bf25}, 874 (1983)].

\end{references}
\end{document}